\documentclass{iau}
\usepackage{graphicx}

\title[Partial Paschen-Back splitting of silicon lines] 
{Partial Paschen-Back splitting of Si\,{\sc ii} and Si\,{\sc iii} lines in magnetic CP stars\footnote{Based on observations obtained at the Canada-France-Hawaii Telescope (CFHT)
       which is operated by the National Research Council of Canada, the Institut National des Sciences
       de l'Univers of the Centre National de la Recherche Scientique of France, and the University of Hawaii.}}

\author[Viktor Khalack \& John Landstreet]   
{Viktor Khalack$^1$
 \and John Landstreet$^{2, 3}$}

\affiliation{$^1$Universit\'{e} de Moncton, Moncton, N.-B., Canada, email: {\tt khalakv@umoncton.ca}\\
$^2$University of Western Ontario, London, Canada \\
$^3$Armagh Observatory, Armagh, Northern Ireland -- United Kingdom}

\pubyear{2013}
\volume{302}  
\pagerange{119--126}
\jname{Title of your IAU Symposium}
\editors{A.C. Editor, B.D. Editor \& C.E. Editor, eds.}
\begin{document}

\maketitle

\begin{abstract}
A number of prominent spectral lines in the spectra of magnetic A and B main sequence stars are produced by closely spaced doublets or triplets. Depending on the strength and orientation of magnetic field, the PPB magnetic splitting can result in the Stokes $I$ profiles of a spectral line that differ significantly from those predicted by the theory of Zeeman effect. Such lines should be treated using the theory of the partial Paschen-Back (PPB) effect. To estimate the error introduced by the use of the Zeeman approximation, numerical simulations have been performed for Si\,{\sc ii} and Si\,{\sc iii} lines assuming an oblique rotator model.
The analysis indicates that for high precision studies of some spectral lines the PPB approach should be used if the field strength at the magnetic poles is $B_{\rm p}>$ 6-10 kG and $V \sin{i} <$ 15 km s$^{-1}$. In the case of the Si\,{\sc ii} line 5041\AA, the difference between the simulated PPB and Zeeman profiles is caused by a significant contribution from a so called ``ghost" line. The Stokes $I$ and $V$ profiles of this particular line simulated in the PPB regime provide a significantly better fit to the observed profiles in the spectrum of the magnetic Ap star HD~318107 than the profiles calculated assuming the Zeeman effect.
\keywords{atomic processes -- magnetic fields -- line: profiles -- stars: chemically peculiar -- stars: magnetic fields -- stars: individual: HD318107}
\end{abstract}

\firstsection 
\section{Introduction}

In the analysis and modelling of spectral lines observed in magnetic upper main sequence stars, it is normally assumed that the splitting of the lines is correctly described by the anomalous Zeeman effect. However, in some cases the fine structure splitting of one or both levels involved in a transition is very small, and magnetic fields found in some stars are large enough to produce magnetic splitting comparable in size to this small fine structure splitting. In these cases, the splitting of the line should be calculated taking into account both the fine structure splitting and the magnetic splitting simultaneously. This regime is known as the incomplete or partial  Paschen-Back (PPB) effect (\cite{P+B21}). The partial Paschen-Back regime can occur when observed spectral line profiles are created by closely spaced doublets or triplets. 


\begin{figure}[t]
\begin{center}
 \includegraphics[width=2.5in, angle=-90]{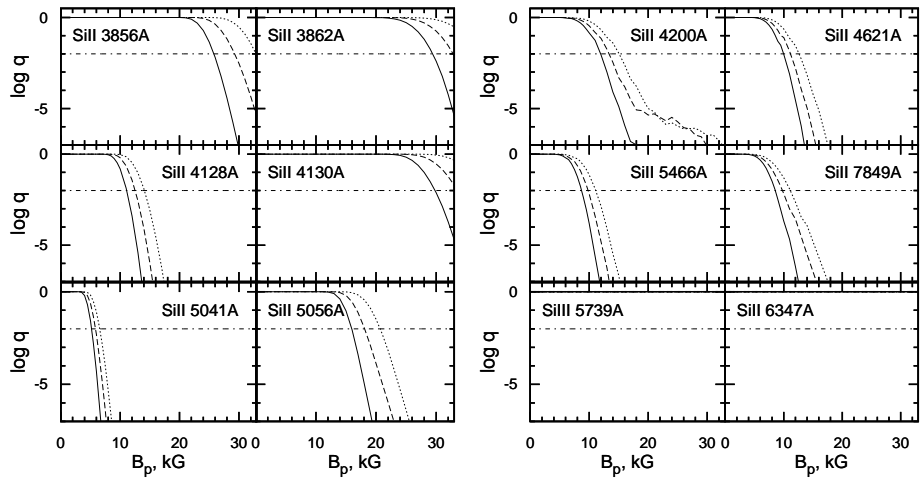}
 \caption{Logarithm of the probability that the noise with $\sigma=0.04$ (S/N=250) is masking the difference between the Stokes $I$ calculated assuming PPB and Zeeman splitting, $\log{[N_{\rm Si}/N_{\rm H}]}=-3.5$ and $V \sin{i}$ = 1 km s$^{-1}$. The continuous, dashed, and dotted lines correspond to the cases where the axis
of magnetic dipole forms an angle $\alpha$= 0$^{\circ}$, 45$^{\circ}$, and 90$^{\circ}$ with the line of sight, respectively.
The horizontal dash-dotted line corresponds to the probability $q$=0.01.
The bottom of the right-hand panel presents the examples of Si\,{\sc ii} and Si\,{\sc iii} lines that show $q\approx1$ for Stokes $I$ profiles with little or no dependence on the field strength.}
   \label{fig1}
\end{center}
\end{figure}

\section{Comparison of PPB and Zeeman profiles}

Calculation of the Paschen-Back splitting of spectral lines has been incorporated into the ZEEMAN2 code (\cite{Khalack+Landstreet12}), which allows us to simulate a line profile composed of several spectral lines (blends), some of which are split in the Paschen-Back regime, while the others are split in the Zeeman regime. Initially, this code was created by \cite{Landstreet88} for the simulation of polarimetric (Stokes $IVQU$ ) line profiles, and was later modified by \cite{Khalack+Wade06}, who added an automatic minimization of the model parameters using the downhill simplex method. 
The procedure for calculation of Paschen-Back splitting takes into account the magnetically perturbed energy levels and determines the respective air wavelength and oscillator strength of components, based on the term configurations and the total strength of all lines in the multiplet under consideration.

The simulation of Stokes $I$ profiles of Si\,{\sc ii} and Si\,{\sc iii} lines is carried for a star with $T_{\rm eff}$=13000K, $\log{g}$=4.0, zero microturbulent velocity, and an oblique rotator model with a dipolar magnetic field structure, assuming that the field strength at the magnetic pole is $B_{p}$ and the the axis of magnetic dipole forms an angle $\alpha$= 0$^{\circ}$, 45$^{\circ}$ and 90$^{\circ}$ with the line of sight. The observed Stokes $I$ spectra are usually contaminated by observational noise $\sigma \approx \frac{1}{S/N}$. To decide whether the difference between the simulated PPB and Zeeman profiles can be confidently detected above the given noise level we use the {\it chi square probability function} (\cite{Abramowitz+Stegun72})
\begin{equation}\label{eq1}
 q(\chi^2|\nu) = 1-p(\chi^2|\nu) = \left[\Gamma\left(\frac{\nu}{2}\right)\right]^{-1} \int_{\chi^2/2}^{\infty}{t^{\frac{\nu}{2}-1} e^{-t}\; dt}
\end{equation}

\noindent where $\nu$ is the number of resolved elements in the analyzed profile and
\begin{equation}\label{eq2}
\chi^2 = \frac{1}{\nu\sigma^2} \sum^{\nu}_{i=1}  (I^{PB}_i - I^{Ze}_i)^2 \;,
\end{equation}

\noindent where $I^{\rm PB}_i$ and  $I^{\rm Ze}_i$ represent the intensity of the Stokes $I$ profiles at a wavelength point $i$ calculated with the assumption of the PPB and Zeeman splitting, respectively. For $\chi^2=35.7$ and $\nu=20$ the probability that PPB and Zeeman profiles are indistinguishable is $q<0.01$. This confidence level seems to be quite robust to perform an evaluation of the difference between the PPB and Zeeman profiles for individual lines (see Fig.~\ref{fig1}).

The Si\,{\sc ii} 5957\AA, 5979\AA, 6347\AA, 6371\AA\, and Si\,{\sc iii} 4552\AA, 4567\AA, 4574\AA\, lines provide $\log q$ close to zero with almost no dependence on the magnetic field strength for the Stokes $I$ profiles. For the Si\,{\sc ii} 3856\AA, 3862\AA\, and 4130\AA\, lines there is no dependence of $\log q$ on magnetic field strength for $B_{\rm p}<$20 kG. Meanwhile, for the Si\,{\sc ii} 4128\AA, 4200\AA, 4621\AA, 5041\AA, 5466\AA\, and 7849\AA\, lines, the simulated PPB and Zeeman profiles appear to be different ($q<0.01$ assuming S/N=250 and $V\sin{i}$ = 1 km s$^{-1}$) for a magnetic field strength $B_{\rm p}$=5-15 kG.

We have also studied the dependence of the probability (see Eq.~\ref{eq1}) on the magnetic field strength for higher rotational velocities (1 km s$^{-1} <V\sin{i}<$ 30 km s$^{-1}$).
The analysis of the Si\,{\sc ii} 4128\AA, 5041\AA, 5466\AA\, and 7849\AA\, lines with S/N = 250 shows that the application of the PPB splitting for the simulation of Stokes $I$ profiles remains important in stars with $V\sin{i}<$ 15 km s$^{-1}$ when the field strength is $B_{\rm p}>$~10-15 kG.

\begin{figure*}[t]
\begin{center}
 \includegraphics[width=1.6in, angle=-90]{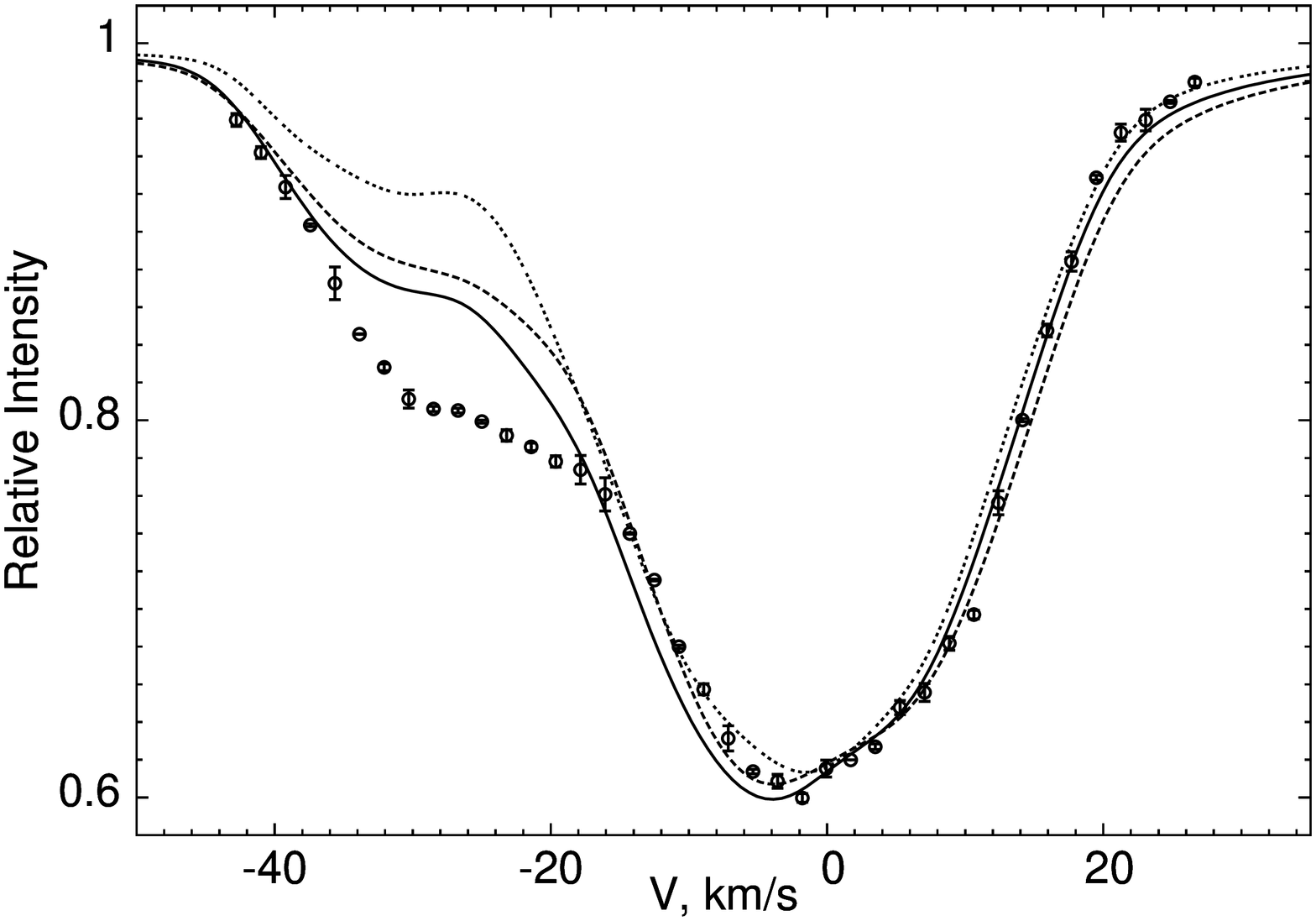}
 \includegraphics[width=1.6in, angle=-90]{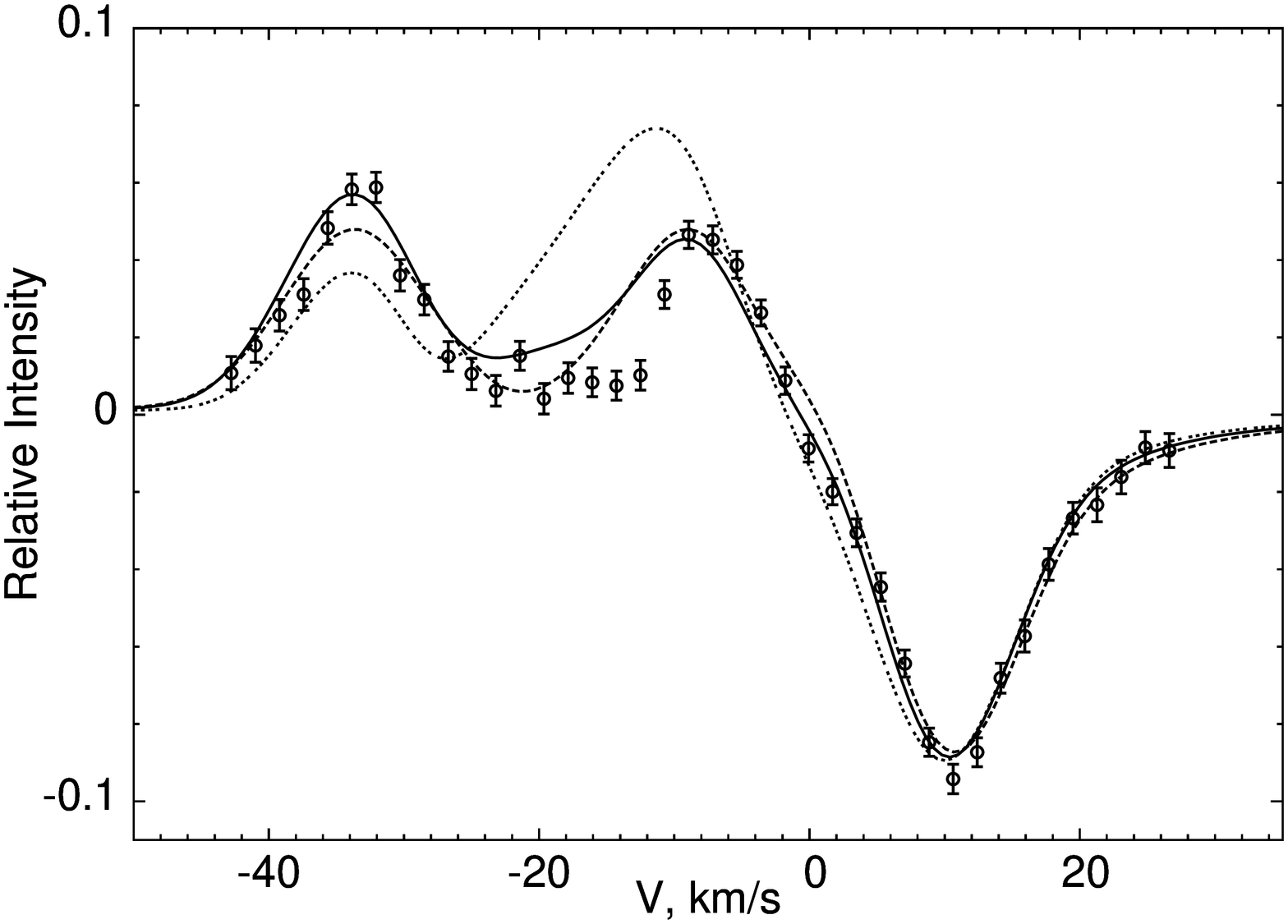}
 \caption{The best fit model for the Stokes $I$ (left) and Stokes $V$ (right) profiles of Si\,{\sc ii} 5041\AA\, line in the Zeeman regime (dotted line) and PPB regime (continuous line) in combination with the Fe\,{\sc i} and Fe\,{\sc ii} blends. Open circles define the observed Stokes I and V profiles of Si\,{\sc ii} 5041\AA\, line and the dashed curve provides the best fit results for this line assuming  PPB splitting without Fe blends. The results for the Zeeman splitting of Si\,{\sc ii} 5041\AA\, line without the contribution from the iron lines are not shown here. }
   \label{fig2}
\end{center}
\end{figure*}

\begin{table}[t]
\centering
\caption[]{Approximation of Si\,{\sc ii} 5041\AA\, line observed in the spectrum of HD~318107 (phase 0.991) employing the PPB and Zeeman effects (see text for details).}
\begin{tabular}{cccccr}
\hline
$\lambda$, \AA& Splitting& $\log[N_{\rm Si}/N_{\rm H}]$ & $V \sin(i)$, km s$^{-1}$ & $V_r$, km s$^{-1}$& $\chi^2/\nu$ \\
\hline
5041& PPB    & -3.57$\pm$0.15& 7.9$\pm$1.0 & -8.4$\pm$1.0 & 16.5  \\
5041& PPB+Fe & -3.80$\pm$0.15& 7.3$\pm$1.5 & -8.1$\pm$1.0 & 9.8   \\
5041& Zeeman & -3.74$\pm$0.15& 7.5$\pm$0.8 & -8.0$\pm$1.0 & 70.7  \\
5041& Zeeman+Fe & -3.94$\pm$0.15& 7.5$\pm$0.8 & -7.7$\pm$1.0 & 40.4  \\
%
\hline
\label{tab1}\end{tabular}
\end{table}

From the bottom of the left-hand panel of Fig.~\ref{fig1} we can see that the PPB effect becomes important for simulations of the Stokes $I$ profile of Si\,{\sc ii} 5041\AA\, line when $B_{p}>$5 kG.
For this line the difference between the simulated PPB and Zeeman profiles appears due to the so called ``ghost" line with $|\Delta J| = 2$ (\cite{Khalack+Landstreet12}). To test the obtained theoretical results, the Stokes I and V profiles of this particular line were analysed for the magnetic Ap star HD318107 ($T_{\rm eff}$= 11800 K, $\log{g}$= 4.2) using both Zeeman and PPB splitting. The geometry of the magnetic field model used for this star is described by the following parameters: $i=22^{\circ}, \beta=65^{\circ}, B_{\rm d}$= 25.6 kG, $B_{\rm q}$= -12.8 kG, $B_{\rm o}$= 0.9 kG that we have adopted from \cite{Bailey+11}. The weak blending lines Fe\,{\sc i} 5040.85\AA, 5040.90\AA\, and Fe\,{\sc ii} 5040.76\AA\, are also taken into account during the simulation, assuming their Zeeman magnetic splitting. From Fig.~\ref{fig2} one can see that the best fit of the observed data is obtained assuming PPB splitting of Si\,{\sc ii} 5041\AA\, line and the contribution of the iron blends (see second line in the Tabl.~\ref{tab1}). The obtained data are close to the results derived by \cite{Bailey+11} from a complex analysis of different spectral lines for this star. The remaining differences between the PPB profiles and the observed spectra may be partially explained in terms of a more complicated actual magnetic field structure and/or horizontal and vertical stratification of the silicon abundance.

\section{Summary}

If the available polarimetric spectra of the stars with strong magnetic field have S/N ratio higher than 250, the use of PPB splitting during the analysis of spectral lines is necessary to obtain precise results in the framework of an assumed model for the abundance map and the magnetic field structure.
In particular, Stokes $I$ 
profiles of some Si\,{\sc ii} lines, when calculated with the PPB splitting, differ significantly from those calculated with the Zeeman effect (see Fig.~\ref{fig1}). This difference appears due to the different relative intensities and positions of split $\sigma$ and $\pi$-components in the PPB and Zeeman regimes, and due to the ``ghost" lines ($|\Delta J| \geq 2$) as in the case of Si\,{\sc ii} 5041\AA\, line (\cite{Khalack+Landstreet12}). For this line profile a contribution from the ``ghost" lines becomes significant for $B_{\rm p}>$~5 kG and an enhanced (by 1 dex) silicon abundance.

\begin{discussion}

\discuss{Romanyuk}{Can you say what the difference will be between the estimates of the magnetic field strength
using the Zeeman and Paschen-Back splitting of magnetically sensitive spectral lines?}

\discuss{Khalack}{This difference depends on the list of spectral lines employed to determine the strength of magnetic field.
If many of those lines appear due to a transition between energy levels characterized by the quantum number $S=0$, the difference should not be large.}

\discuss{Wade}{You have mentioned the significant difference between the Stokes I (or Stokes V) profiles simulated using the Zeeman and Paschen-Back splitting.
What can you say about the Stokes Q and U profiles?}

\discuss{Khalack}{A comparison of the simulated profiles has been done under the assumption of S/N=250 for Stokes I spectra and a homogeneous distribution of silicon abundance.
For the analysed silicon lines and field strength at the magnetic pole $B_{\rm p} < 30 kG$, the Stokes Q and U profiles appear to be rather weak (comparable to noise) and no significant difference can be detected in this case.}

\end{discussion}

\end{document}